\documentclass{kluwer}
\usepackage{times}
\usepackage{graphics}
\usepackage{graphicx}
\usepackage{epsf}
\usepackage{epsfig}

\newcommand{\be}{\begin{equation}}
\newcommand{\ee}{\end{equation}}
\newcommand{\BE}{\begin{eqnarray}}
\newcommand{\EE}{\end{eqnarray}}
\newcommand{\BEn}{\begin{eqnarray*}}
\newcommand{\EEn}{\end{eqnarray*}}
\newcommand{\barr}{\begin{array}} 
\newcommand{\earr}{\end{array}}
\newcommand{\bit}{\begin{itemize}}      
\newcommand{\eit}{\end{itemize}}
\newcommand{\bfl}{\begin{flusleft}}
\newcommand{\efl}{\end{flusleft}}
\newcommand{\bfr}{\begin{flushright}}
\newcommand{\bc}{\begin{center}}
\newcommand{\ec}{\end{center}}
\newcommand{\ben}{\begin{enumerate}}    
\newcommand{\een}{\end{enumerate}}
\newcommand{\cl}{\centerline}

\renewcommand{\theparagraph}{\Alph{paragraph}}

\begin{document}
\begin{article}
\begin{opening}
\title{A dynamical approach to protein folding\\}

\author{Alessandro \surname{Torcini}\thanks{E-mail: 
torcini@ino.it - URL: htpp://www.ino.it/
$\sim$torcini}} 
\institute{Dipartimento di Fisica, Universit\'a "La Sapienza",
P.zle A. Moro, 2 - I-00185 Roma, Italy\\
INFM, UdR Firenze, L.go E. Fermi, 2 - I-50125 Firenze, Italy}
\author{Roberto \surname{Livi}\thanks{E-mail: livi@fi.infn.it}}
\institute{Dipartimento di Fisica, Universit\'a di Firenze, 
        L.go E. Fermi, 2 - I-50125 Firenze, Italy\\
        INFM, UdR Firenze,
        L.go E. Fermi, 2 - I-50125 Firenze, Italy}
\author{Antonio \surname{Politi}\thanks{E-mail: politi@ino.it}}
\institute{Istituto Nazionale di Ottica Applicata,
        L.go E. Fermi, 6 - I-50125 Firenze, Italy\\
        INFM, UdR Firenze,
        L.go E. Fermi, 2 - I-50125 Firenze, Italy}

\runningtitle{A dynamical approach to protein folding}
\runningauthor{A. Torcini, R. Livi, \& A. Politi}

\dedication{To Linus Torvalds}

\begin{motto}
What can be thought, can be simulated.
\end{motto}

\begin{abstract}
In this paper we show that a dynamical description of the protein folding 
process provides an effective representation of equilibrium properties 
and it allows for a direct investigation of the mechanisms 
ruling the approach towards the native configuration. 
The results reported in this paper have been obtained for
a two-dimensional toy-model of amino acid sequences, whose
native configurations were previously determined by
Monte Carlo techniques.
The somewhat controversial scenario emerging from 
the comparison among various thermodynamical
indicators is definitely better resolved relying upon
a truly dynamical description, that points out the crucial
role played by long-range interactions in determining
the characteristic step-wise evolution of ``good'' folders to their native 
state. It is worth stressing that this dynamical scenario is consistent
with the information obtained by exploring the energy
landscapes of different sequences. This suggests that even the identification
of more efficient ``static'' indicators should take into account the peculiar
features associated with the complex ``orography'' of the landscape.
\end{abstract}


\end{opening}

\section{Introduction}

\noindent
Proteins are heteropolymer chains made of aminoacids. The aminoacid sequence
(the so-called primary structure) determines the native configuration 
(tertiary structure) which, in turn, is responsible for the biological 
activity of the protein. The identification of the native structure 
corresponding to a given aminoacid sequence and, viceversa, of the sequence 
yielding a given configuration are called direct and inverse problem, 
respectively. In spite of the increasing efforts made 
by the researchers working in this area, both problems remain generally
unsolved. A few different strategies have been adopted so far by the 
scientific community to tackle the protein-folding problem. The first method
that has been developed could be called ``black-box'' approach, since 
one tries to infer the tertiary structure with no other knowledge than 
the configurations corresponding to some specific aminoacid sequences
(e.g., the neural-network approach). Although this method has been
implemented with some success, the lack of information about the physics
of the underlying folding process does not allow going beyond statistical 
predictions. In order to overcome such difficulties, simplified Hamiltonians 
have been introduced with the goal of identifying the native structure through
the implementation of equilibrium-statistical-mechanics tools (e.g. Monte
Carlo techniques). The main difficulty of this approach arised from the huge 
number of relative minima, which makes the search for the absolute minimum 
rather questionable in realistic cases. 

However, it is known that, in spite of the very many accessible configurations,
the protein folding turns out to be rather fast, actually, much faster
than a pure random search (once the appropriate time scales are taken
into account) \cite{crei}.
It is, therefore, rather tempting to tackle the problem from a pure
dynamical point of view, following, e.g., the evolution of ``coiled'' 
configurations towards globular-folded structures. An ``ab initio'' approach, 
where all molecular forces acting among the protein elements and between 
protein and solvent are taken into account, should, in 
principle, reveal all details of the folding dynamics. Unfortunately, even 
if the degrees of freedom of the solvent are traced out from the interaction 
Hamiltonian, the characteristic times associated with the microscopic dynamics 
are on the order of ${\cal O}(10^{-11})$ seconds, while the folding process
is expected to occur typically on time scales in between ${\cal O}(10^{-2})$ 
and ${\cal O}(1)$ seconds. Simulating systems with thousands of degrees of 
freedom over time scales that cover ten orders of magnitude is definitely out 
of reach for the actual computing facilities and it will remain as such at 
least in the near future. 

On the other hand, it appears reasonable to conjecture that the fine details 
regarding the interaction structure and the degrees of freedom corresponding
to the inner dynamics of the aminoacids do not matter for the folding process. 
Therefore, one can employ ``coarse grained'' potentials, epitomizing only a 
few relevant interactions. The price payed for such a drastic 
reduction of the gigantic complexity of the molecular structure of a protein 
should be hopefully compensated by the possibility of obtaining a reliable 
description of the folding process (provided the main ingredients ruling such 
a process have been correctly identified).

In fact, it seems that evolution has selected proteins out of all possible 
aminoacid sequences in such a way that their native states are stable and 
kinetically accessible, so that only those sequences satisfying both 
requirements are biologically active. In fact, a great deal of papers has 
been devoted to the attempt of identifying ``bad" and ``good" folder 
sequences, relying upon their structural or equilibrium properties
\cite{thir1,thir2,shak,shak1,irback1,irback2}.

With reference to a 2D off-lattice model, in this paper we show that strictly 
dynamical simulations can provide a full acount of heteropolymer properties. 
In particular, equilibrium simulations allow for an effective
identification of the lowest minima of the energy landscape. Moreover, the
comparison between folding and unfolding simulations shed some light on
the glassy transition, while the peak of the specific heat is clearly
resolved to locate the collapse transition. Throughout the paper we
compare the behaviour and the properties of five sequences, suitably selected
to investigate the differences between possible ``good'' and ``bad''
folders.

More specifically, in section II we introduce the mesoscopic 2D
off-lattice model. Equilibrium thermodynamic properties of the five
selected sequences are discussed in section III (looking both at
standard observables such as the total energy and the average distance
between configurations). The conformation of the native valley and the 
associated energy funnel are investigated in section IV, while section V is 
devoted to the description of the dynamical evolution. Concluding remarks 
are reported in Sec. VI.

\section{The Model}

We will consider a slight generalization of the 2-dimensional off-lattice model
recently introduced by Stillinger {\it et al.} \cite{still} and similar to
that one previously studied by Iori et al. \cite{iori}. Such a model is
characterized by $L$ point-like monomers (mimicking the residues of a
heteropolymer) arranged along a one dimensional chain. The nature of the
residues is assumed for simplicity to be of two types only: hydrophobic (H) or 
polar (P). Thus, each heteropolymer is unambiguously identified by a sequence
of binary variables $\{ \xi_i \}$ (with $i=1, \dots, L$) along the backbone,
where $\xi_i = 1$ if the $i$th residue is of type H and $\xi_i = -1$,
otherwise. The intramolecular potential is composed of three terms for each 
monomer: a nearest-neighbour harmonic interaction ($V_1$), a three-body 
interaction ($V_2$) to simulate the energy cost of local bending, and a 
Lennard-Jones--like (LJ) interaction ($V_3$) acting between pairs $(i,j)$ of 
non-neighbouring residues. This last term depends on the nature of the 
residues, i.e. on both $\xi_i$ and $\xi_j$, in such a way to mimic the
interaction with the solvent.

The Hamiltonian of the system writes as
\begin{equation}
H = \sum_{i=1}^L \frac{p_{x,i}^2+p_{y,i}^2}{2} +
\sum_{i=1}^{L-1} V_1(r_{i,i+1}) + \sum_{i=2}^{L-1} V_2(\theta_i)
+ \sum_{i=1}^{L-2} \sum_{j=i+2}^{L}  V_3(r_{ij},\xi_i,\xi_j)
\label{hamil}
\end{equation}
where the mass of each monomer is assumed to be unitary,
$(p_{x,i},p_{y,i}) = ({\dot x}_i,{\dot y}_i)$, and
$r_{i,j}=\sqrt{(x_i-x_j)^2+(y_i-y_j)^2}$.
The first potential term appearing in Eq.~(\ref{hamil}) is
\begin{equation}
V_1 (r_{i,i+1}) = \alpha (r_{i,{i+1}}-r_0)^2
\label{v1}
\end{equation}
with $\alpha=20$ and $r_0 = 1$; the second term, favouring the chain
alignment, reads 
\begin{equation}
V_2(\theta_i) = \frac{1 - \cos \theta_i}{16}
\label{v2}
\end{equation}
where
\begin{equation}
\cos \theta_i = \frac{ (x_i -x_{i-1})(x_{i+1}-x_i) +
(y_i -y_{i-1})(y_{i+1}-y_i)}{r_{i,i-1}r_{i+1,i}}
\label{coste}
\end{equation}
and $-\pi < \theta_i < \pi$.
The last, nonlocal, interaction is
\begin{equation}
V_3(r_{i,j}) = \frac{1}{r_{i,j}^{12}} - \frac{c_{i,j}}{r_{i,j}^6}
\label{v3}
\end{equation}
where $|i-j| >1$ and 
\begin{displaymath}
c_{i,j} = \frac{1}{8} (1+\xi_i + \xi_j +5 \xi_i \xi_j) 
\quad   .
\end{displaymath}
Accordingly, the interaction is attractive if both residues are either 
hydrophobic or polar (since $c_{i,j} = 1$ and $1/2$, respectively), 
while it is repulsive if the residues belong to different species 
($c_{ij} = -1/2$).  The only difference with the model introduced by Stillinger 
{\it et al.} \cite{still} comes from the nearest-neighbour interaction: 
the originally rigid bond is here replaced by the harmonic term $V_1$.
We have preferred this latter choice, because it represents a more
realistic nearest neighbours interaction. Anyhow, the large value of the 
coupling constant $\alpha$ herein adopted makes the difference rather irrelevant.

Quite accurate Monte-Carlo (MC) simulations, performed by employing innovative 
schemes, have revealed that, analogously to real proteins, only a few 
sequences fold into a native structure (good folders), while the majority of 
the possible sequences do not possess a unique folded state 
\cite{irback1,irback2}.

The dynamics of the toy model (\ref{hamil}) has been investigated by 
integrating the corresponding Hamilton-Jacobi equations in the presence of a 
heat bath. The thermal reservoir has been simulated by separately implementing 
a Nos\'e-Hoover thermostat for each residue of the chain, while the integration 
has been performed by employing a second order Runge-Kutta scheme. The 
evolution equations read
\begin{eqnarray}
&& {\dot x}_i = p_{x,i} \quad ; \quad {\dot y}_i = p_{y,i}\\
&& {\dot p}_{x,i} = - \frac{\partial H}{\partial x_i} - \zeta_i p_{x,i} \quad ;
\quad 
{\dot p}_{y,i} = - \frac{\partial H}{\partial y_i} - \zeta_i p_{y,i} 
\\
&& {\dot \zeta}_i = \frac{1}{\tau^2} \left(
\frac{p_{x,i}^2 + p_{y,i}^2}{2 T_b} - 1 \right)
\label{hj}
\end{eqnarray}
where $\zeta_i$ represents the ``bath'' variable that acts to keep the 
temperature of the $i$th residue at the constant value $T_b$, and $\tau$ is 
the ``reaction'' time 
of the bath (typically set equal to 1 in our simulations). Numerical 
integrations have been performed with a time-step $\delta t = 0.025$~, after 
having verified that this value is small enough to guarantee a good accuracy.

Two different kinds of dynamical simulations have been performed, namely 
unfolding (US) and folding (FS) simulations. In the first case, the initial 
state of the ``protein'' is taken equal to the native configuration (NC), that 
we assume to coincide with the minimal energy configuration.
Thermodynamic quantities have been thereby determined by averaging 
fixed-temperature simulations over a time interval $t \sim 5 \cdot 10^5$. 
FS's have instead been performed starting from an initial configuration 
generated by setting the residues at a fixed distance $r_{i,j}=r_0$ with 
randomly 
distributed angles $\theta_i$ within the interval $[-\pi/4;\pi/4]$. The system 
is then let relax for a time $t_{r}$ that has been fixed depending on the
simulation temperature (from $t_r \sim 5 \cdot 10^5$ to $t_r \sim 10^7$ in the 
temperature range considered later on). After this transient, the various
observables have been averaged over a time interval ranging from 
$5 \cdot 10^5$ to $1.9 \cdot 10^6$. Additionally, we have averaged over 10 
different initial conditions.

In order to investigate the folding properties of this toy model, we have 
studied a homopolymer of length 20 and 4 heteropolymers each composed of 
14 H-type 
and 6 P-type residues. To be more specific, we have analyzed the dynamical 
and thermodynamical properties of the following five sequences :
\begin{itemize}

\item{[S0]} a homopolymer composed of hydrophobic residues (i.e.,  
$\xi_i=-1, i=1,\dots,L$);

\item{[S1]=[HHHP HHHP HHHP PHHP PHHH]} 
a sequence previously analyzed in \cite{irback2}, where it
was identified by the code number 81 and recognized as a good folder for 
the Stillinger model \cite{still}; 

\item{[S2]=[HHHH PHHP HPHP HHHH PHPH]} 
the sequence with the maximal $Z$-score \cite{bowie,mirny} within an
ensemble of 6,900 sequences each composed of 14 H- and 6 P-type
residues \footnote{The definition of the $Z$-score is
$$
Z = (V_{NC} - \langle V \rangle)/ W
$$
where $V_{NC}$ is the potential energy of the NC, $\langle V\rangle$ is the 
average potential energy of a suitable set of alternative configurations and
$W =\sqrt{\langle V^2 \rangle-\langle V \rangle ^2}$. In order to select such 
configurations, we have first identified $467$ distinct inherent minima 
(see Sec.~4 for the definition) of the homopolymer. These minima have been
then considered as the initial condition for a gradient method to identify
the closest local minimum for each sequence. For the sequence S2, $Z=-5.70$,
while for S1, $Z=-2.97$ (notice that in both cases the NC does not belong to 
the set of alternative configurations over which the average has been 
performed).

Moreover, for each of the five sequences studied in this paper, the $Z$-score 
has been evaluated also identifying the ``alternative'' configurations with 
the inherent minima as determined from simulations performed at a temperature 
$T=0.08$. In this case, the values are $Z=-4.50$ [S1], $Z=-3.16$
[S3], $Z=-3.08$ [S4], $Z=-2.98$ [S2] and $Z=-2.20$ [S0]. Accoridngly,
the best folder seems to be S1, while the worst one is S0.};

\item{[S3]=[PHPH HHHH HHPH HHHHP HHPP]} 
a sequence identified by the code number 50 in Ref.~\cite{irback2},
where it was recognized as a bad folder;

\item{[S4]=[PPPH HPHH HHHH HHHP HHPH]} 
a randomly generated sequence of 14 H- and 6 P-type residues.

\end{itemize}

\section{Equilibrium Properties}

\subsection{Standard thermodynamic observables}

Before investigating the protein-like properties of the heteropolymer dynamics,
we have investigated standard  equilibrium-thermodynamics observables. Let us
start defining the temperature as
\begin{equation}
T = \frac{1}{L} \left< \sum_{i=1}^L \frac{p_{x,i}^2+p_{y,i}^2}{2} \right> ,
\label{temp}
\end{equation}
where the Boltzmann constant has been set equal to one, while 
$\langle \cdot \rangle $ 
denotes a time average along the trajectory in the phase space (notice that
the thermal baths defined in the previous section induce a
canonical-ensemble measure in the phase space).

\begin{figure}[ht]
\cl{
\epsfig{figure=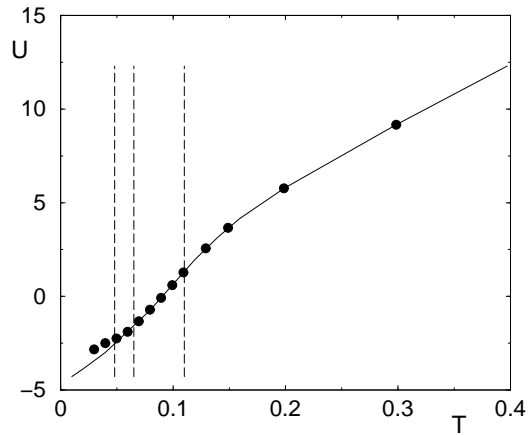,angle=-90,width=7truecm}
}
\caption{The total energy $U(T)$ in equilibrium simulations for the sequence 
S1: the solid line corresponds to US's, while the symbols refer to FS's. 
The dashed lines correspond to (from left to right), $T_G$, $T_F$, and 
$T_\theta$ (notice that for this sequence, $T_\theta \approx
T_\theta^*$).}
\label{f3}
\end{figure}

In all cases, at sufficiently large temperatures, the averages obtained 
from US's and FS's do coincide: this indicates that the time span of the 
simulations is long enough to guarantee a good equilibration of
the measure. At lower temperatures, the heteropolymer structure can be trapped
in local minima of the potential, thus yielding different results for the 
finite-time US's and FS's~. This is 
illustrated in Fig.~\ref{f3}, where we have reported $U(T)$ for the sequence 
S1. Although the difference between US's and FS's depends on the time
span used in the averages, the expected exponential growth of the time needed 
to visit ergodically the whole phase-space makes it sensible to introduce
a rough definition of ``glassy'' temperature, $T_G$, 
as the temperature below which the relative 
difference between the values of the internal energy estimated with the 
two procedures is larger than $10\%$. 

The slope of $U(T)$, i.e. the specific heat $C_V$, exhibits a clear peak at
a temperature $T_\theta^* \approx 0.1$~. Although one cannot speak of 
phase-transitions in finite systems, this behaviour is definitely reminiscent 
of the $\theta$-transition firstly studied in homopolymers \cite{degennes},
where a low-temperature phase, characterized by compact configurations, and 
a high-temperature phase, characterized by random-coil states, have been 
identified. In the context of protein-like chains, this translates into the
so-called {\it collapse transition} \cite{thir1,thir2}, that is identified 
as the temperature corresponding to the maximum of $C_V$. Moreover, 
this $U(T)$ dependence is also peculiar of systems with attractive
interactions, where the collapse transition occurs when such interactions 
become dominant over the other energy contributions (see, e.g., 
self-gravitating systems, atomic and molecular clusters) \cite{ar,ta,haberland}.

\begin{figure}[ht]
\cl{
\epsfig{figure=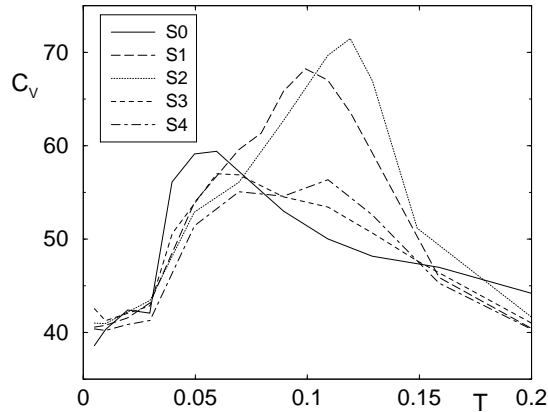,angle=-90,width=7truecm}
}
\caption{The specific heat $C_V$ as a function of $T$. All data refers 
to US's.}
\label{f4}
\end{figure}

In practice, the specific heat is better estimated by looking at the 
fluctuations of the internal energy, 
\begin{equation}
C_V = \frac{ \langle U^2 \rangle - \langle U \rangle^2}{T^2} .
\label{cv}
\end{equation}
The numerical results obtained for the sequences S0-S4 are reported in
Fig.~\ref{f4}. There we see that all curves start from $C_V \simeq 40$ 
to exhibit a more or less broad peak. In fact, at sufficiently
low-temperatures, any sequence is practically indistinguishable from a
(disordered) 2d solid, in which case the specific heat is equal to $2L$.
At high temperatures, $C_V$ seems to converge to some lower value.
In a generic chain with nearest-neighbour interactions 
in a plane we expect a $T/2$ contribution from each one of the kinetic 
degrees of freedom and only one  $T/2$ contribution from the components
of the potential energy, dominated by the longitudinal interaction. 
Altogether this implies that the specific heat should be 
$C_V \simeq 3L/2 = 30$. In practice, we find slightly larger values, as 
shown in Fig.~\ref{f4}. 
The difference has to be attributed to the 
Lennard-Jones potentials that are not yet completely negligible at 
temperatures close to $0.4$ (the energy contribution of the angular 
term turns out to be fairly independent of $T$). In fact, the interparticle 
distance grows with the temperature and, accordingly, the LJ energy 
increases as well, converging to 0 from below. 

Moreover, we notice that S1 and S2 exhibit both the largest collapse 
temperatures and the highest $C_V$-peaks. For the other three sequences
$T_\theta^*$ is smaller by approximately a factor of two, while the peak value 
of $C_V$ is $15 - 20 \%$ lower than those obtained for S1 and S2.
This suggests that maximizing the zeta score is a good strategy at least
for optimizing the collapse transition. 
This can be investigated more directly by looking at the
gyration radius
$$
R_{gy} = \sqrt{ \left \langle \frac{1}{N} \sum_{i=1}^N |{\bf r}_i - 
  {\bf r}_{cm}|^2 \right \rangle } \quad ,
$$
where ${\bf r}_{cm}$ denotes the center of mass of the sequence and the 
average is performed over configurations generated by the dynamical 
evolution of the system. The indication for the collapse transition is
usually associated with a rapid decrease of $R_{gy}(T)$ close to a temperature
$T=T_\theta$, where its derivative should exhibit a singularity. 
However, as already observed in
\cite{irback2}, the finiteness of the chains can at most yield a 
smooth decrease of $R_{gy}(T)$, being the singularity intrinsic to
thermodynamic limit properties.
In finite systems a generally accepted estimate of 
$T_\theta$ is the temperature at which 
$\partial R_{gy}(T)/\partial T$ is maximal.

\begin{figure}[ht]
\cl{
\epsfig{figure=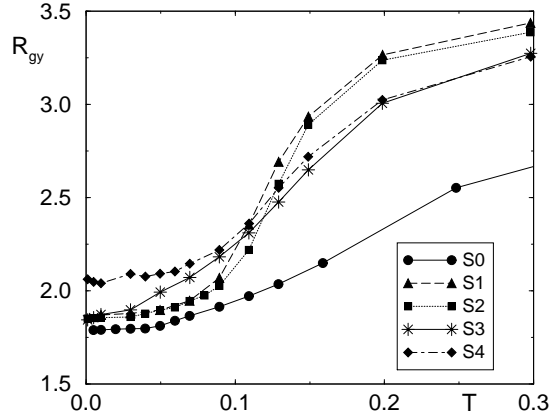,angle=-90,width=7truecm}
}
\caption{Gyration radius as a function of the temperature for all S0-S4 
sequences. The data refers to US's.}
\label{f5}
\end{figure}

With this definition, we obtain for $T_\theta$
essentially the same value of $T_\theta^*$ for the sequences S1 and S2,
while a $T_\theta >> T_\theta^*$ for all the other 3 sequences.
Obviously, the homopolymer turns out to be the most compact sequence at all
temperatures, as all LJ-potentials are attractive. Coherently with the
previous analysis, we can notice a similar behaviour for $S1$ and $S2$ which
again display a more pronounced transition-like behaviour. A further
observation concerns the relatively larger gyration radius exhibited by S4
at low temperature: we can imagine that the frustration of its typical random 
structure prevents the formation of a more compact ``native'' configuration. 

Some of the indicators computed for the S0-S4 sequences are reported in 
Table~I. The largest value for the temperature of the ``glassy'' transition
is obtained for S1 (namely $T_G \simeq 0.048$). Within the framework of the 
random energy model applied to the protein-folding problem, it is commonly 
believed that a good folder is characterized by a large ratio $\rho = T_F/T_G$
\cite{rem} (where $T_F$ is the {\it folding temperature}, defined in 
Subsec. 4.2). From the data reported in Table I, S1 (identified as a good 
folder in \cite{irback2}) should be definitely the worst one! It is not clear 
whether such an akward conclusion implies that $\rho$ is not a meaningful 
indicator as hoped for, or whether the sequence S1 is not a good folder as 
previously believed. 
 
\begin{table}[ht]
\caption[tabone]{
For all S0-S4 sequences we report: the glassy temperature $T_G$; the ratio 
$\rho= T_F/T_G$ (the folding temperature $T_F$ can be found in Table III); 
the collapse-transition temperature as estimated from the gyration radius 
($T_{\theta}$) and from the specific heat ($T_{\theta}^*$); the maximum
value of $C_V$~.}
\vskip 0.3 truecm
\begin{tabular}{rrrrrr}
\hline
\hfil  \hfil \hfil & \hfil S0 \hfil & \hfil S1 \hfil &
\hfil  S2  \hfil & \hfil S3  \hfil & \hfil S4 \hfil \hfil \\
\hline
$T_G$ &  0.022  &  0.048 & 0.028   & 0.038 & 0.025 \\
$\rho$ &  1.63  &  1.36  & 1.71 & $\simeq 0.95$   & 1.84 \\
$T_\theta$ &  0.16  &  0.11 & 0.13   & 0.13 & 0.13 \\
$T_\theta^*$ &  0.05  &  0.105 & 0.12   & 0.06 & 0.06 \\
$C_V(T_\theta^*)$ &  60  & 68 & 72 &  57 & 55.5  \\
\hline
\end{tabular}
\end{table}

\subsection{Distance between confingurations}

\begin{figure}[ht]
\cl{
\epsfig{figure=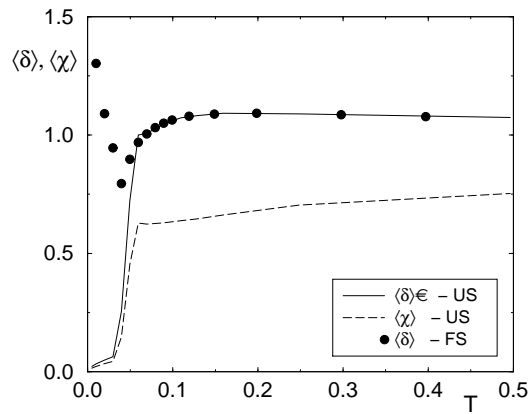,angle=-90,width=7truecm}
}
\caption{
The average distance $\langle \delta\rangle$ for US's (solid line) and 
FS's (circles) as a function of the temperature $T$, reported toghether 
with the average value of the structural overlap function 
$\langle \chi \rangle$ for US's (dashed line) in the homopolymer S0.}
\label{f6}
\end{figure}

\begin{figure}[ht]
\cl{
\epsfig{figure=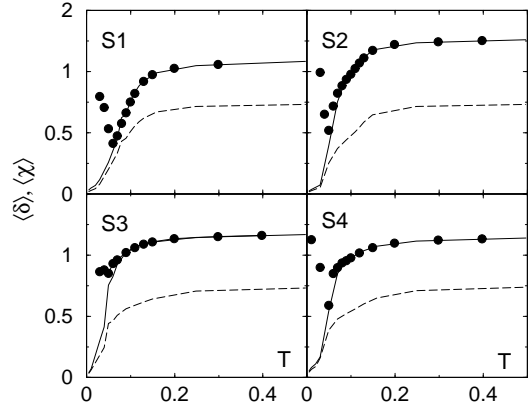,angle=-90,width=7truecm}
}
\caption{
The same quantities as in the previous figure for the sequences S1-S4.}
\label{f7}
\end{figure}

In order to study protein-like features of equilibrium simulations, it
is important to look at the shape of the heteropolymer and, in particular,
to quantify differences between shapes. Given any two configurations
$\alpha$ and $\beta$, identified by the angle sequences 
$\theta^{(\alpha)}_i$, $\theta^{(\beta)}_i$ ($2<i<L-1$, see Eq.~(\ref{coste})~), 
we introduce the following angular distance
\begin{equation}
\delta (\alpha,\beta) := \min \left\{ \frac{1}{L-2} \sum_{i=2}^{L-1} \left |
\theta_i^{(\alpha)}
-\theta_i^{(\beta)}\right | \right\} \quad ;
\label{angdist}
\end{equation}
where the minimum is taken over reflections only, since this distance
is invariant with respect to translations and rotations of any single 
configuration. Tipically, we are interested in looking at the angular 
distance between any dynamical configuration of a sequence and its 
corresponding NC. For the sake of brevity, we indicate this distance
with $\delta$, omitting the explicit dependence on the generic dynamical 
configuration.  In the following, we will show that $\delta$ provides 
essentially the same information as the structural overlap function 
\cite{thir1}
$$
\chi := 1 - \frac{2}{(L-1)(L-2)} \sum_{i=1}^{L-2} \sum_{j=i+2}^L
\Theta(\varepsilon-|r_{ij}-r_{ij}^N|)
\label{chi}
$$
where $r_{ij}$ is the distance between the $i$th and $j$th residues of a given 
configuration, $r_{ij}^N$ is the corresponding distance in the NC,
$\Theta(x)$ is the Heaviside function and $\varepsilon$ 
denotes a suitably chosen threshold. 
In order to compare this indicator with $\delta$ we have
adopted a slightly different, but practically equivalent, definition
by replacing $\Theta(x)$ in Eq.~(\ref{chi}) with 
$(1+\hbox{tanh}(\kappa x))/2$.
It can be readily verified that $1/\kappa$ plays the role of $\varepsilon$;
in our calculations we have set $\kappa = 1$.
The average angular distance $\langle \delta \rangle$ is reported in 
Figs. (\ref{f6}-\ref{f7}) for all S0-S4 sequences with both FS's and US's.
At sufficiently high temperatures ($T > T_G$) the results of FS's and 
US's concide, as the observables introduced in the previous section do. 
Typically, $\langle \delta \rangle $ increases with $T$ and eventually 
saturates to a sequence-dependent asymptotic value $\delta_a$ for $T>0.2$~.
The comparison with the behaviour of the average structural overlap 
$\langle \chi \rangle$ (reported in Figs.~(\ref{f6}-\ref{f7}) for US's) 
indicates that this quantity is practically equivalent
to $\langle \delta \rangle$, the only difference being an irrelvant scale
factor. 

The dependence of $\langle \delta \rangle$ on the temperature $T$, obtained
with the US's, gives a first rough idea of the shape of the native valley. 
The slower growth observed for S1 suggests that this sequence is characterized 
by a wider basin of attraction. Some information about the ``accessibility'' 
of the native valley can, instead, be extracted from the difference between 
$\langle \delta\rangle$ obtained with FS's and US's. If, during the folding 
dynamics, the heteropolymer is unable to enter the native valley in a broad
temperature range (within the employed integration time) this is a strong 
indication that the corresponding sequence is a slow folder.

In order to compare the different sequences on a more quantitative 
level, we have introduced the ``foldability" quality-factor
$$ 
Q := \frac{\delta_a}{\delta_m}
$$
where $\delta_m$ is the minimal value reached by $\langle \delta \rangle$ 
during the FS's.  A high $Q$-value indicates that the protein noticeably 
approaches the native structure before the structural arrest sets in 
below $T=T_G$. Conversely, a relatively small $Q$-value suggests that the 
protein does not even enter the native valley before the dynamics is 
dramatically slowed down at the glassy transition. The data reported in 
Table~II indicate that the largest $Q$-values are obtained for $S1$ and $S_2$,
indicating that the only two ``good'' folders are $S1$ and $S2$.

\begin{table}[ht]
\caption[tabone]{
The foldability factor $Q$, the temperatures $T_F^\chi$, $T_F^\delta$,
$\sigma_\chi$, $\sigma_\delta$, and $\sigma^*$ for the 5 considered sequences.
The values reported within parentheses for S2 correspond to a second peak
shown by $\chi$ for this sequence.
}
\vskip 0.3 truecm
\begin{tabular}{rrrrrr}
\hline
\hfil  \hfil \hfil & \hfil S0 \hfil & \hfil S1 \hfil &
\hfil  S2  \hfil & \hfil S3  \hfil & \hfil S4 \hfil \hfil \\
\hline
$Q$ & 1.37  & 2.63  & 2.44  &  1.37 & 1.92  \\
$T_F^\chi$ & 0.05  & 0.07  & 0.05 (0.12)  &  0.04  & 0.05  \\
$T_F^\delta$ & 0.04  & 0.10  & 0.07   &  0.06  & 0.07  \\
$\sigma_{\chi}$ & 0.09  & 0.33  & 0.58 (0.00)  &  0.33  & 0.55  \\
$\sigma_{\delta}$ & 0.27  & 0.05  & 0.41   &  0.00  & 0.37  \\
$\sigma^*$ & 0.78  & 0.41  & 0.63   &  0.71  & 0.64  \\
\hline
\end{tabular}
\end{table}

\noindent

In \cite{thir1,thir2} it was suggested that the variance of $\chi$ is a
meaningful indicator to define the {\it folding temperature}. More precisely, 
it is claimed that, analogously to the collapse transition identified from 
the maximum of the fluctuations of the internal energy, the folding 
temperature $T_F^{\chi}$ corresponds to the peak of the variance of the
structural overlap $\chi$. The almost equivalence between $\chi$ and the
angular distance $\delta$ suggests that one could equally look at
\begin{equation}
\Delta \delta  = \langle(\delta)^2\rangle - \langle(\delta)\rangle^2
\label{vardist}
\end{equation}
where the average $\langle \cdot \rangle$ is taken over all the possible 
configurations taken by a heteropolymer (with a given sequence) during a 
certain time evolution and over different initial conditions. Accordingly,
the temperature corresponding to the peak of the variance 
$\Delta \delta$ 
should identify the folding temperature. As it is not a priori obivous that
this second definition coincides with the former one, we shall denote it by
$T_F^\delta$. The results of US's for both $T_F^\delta$ and $T_F^\chi$ 
are reported in Table~II (those referring to $T_F^\delta$ are confirmed by 
independent FS's). One can see that they are qualitatively similar, but
not really close to each other. This should be taken as an indication that
the concept of ``folding transition'' is ill-defined as all transitions in
finite systems.

The ``Camacho-Klimov-Thirumalai criterion" states that when the 
parameter
$$
\sigma_\chi = \frac{|T_\theta^* - T_F^\chi|}{T_\theta^*}
$$
is small (e.g., $\le 0.4$ for off-lattice models), the corresponding sequence 
is a fast folder, while slow folders are characterized by $\sigma_\chi > 0.6$ 
\cite{veit,thir2}.
In our case we have estimated both $\sigma_\chi(T)$ and $\sigma_\delta(T)$ 
(with an obvious meaning of the notation) and the corresponding values
are reported in Table~II. As a first observation, notice that $\sigma_\chi(T)$ 
and $\sigma_\delta(T)$ do not give coherent indications. By looking at the
values of $\sigma_\chi$ we are lead to the ``absurd'' conclusion that the
only good folder is the homopolymer and that all other sequences appear as 
not-so-fast folders. On the other hand, by looking at $\sigma_\delta$, one 
deduces that the fast folders are $S_1$ and $S_3$, a conclusion that is 
still partly inconsistent with \cite{irback2}, where it was ascertained that
S3 is a bad folder. Thus, we can only conclude that in our case, the 
Camacho-Klimov-Thirumalai criterion does not help in properly identifying 
good folders. 

\noindent
However, if we replace $T_\theta^*$ with $T_\theta$ (i.e. if we look at the
peak of the fluctuations of the gyration radius) and define $T_F$ as discussed
in the next chapter (see Table~III), 
we obtain the much more meaningful indicator $\sigma^*$ . In fact, 
from the data reported in the last row of Table~II, one concludes that
$S1$ is the only reasonably good folder. Clearly, such differences indicate
that finite-size corrections are too important to be neglected in
this type of studies. It would be really useful to study a much larger 
ensemble of systems to conclude whether $\sigma^*$ is a truly trustful 
indicator.  In any case, it remains to be understood why  $\sigma^*$ is more 
meaningful than other, apparently equivalent, indicators.

\section{Energy Landscape}

\subsection{Native Configurations}

Assuming that the native configuration of each sequence coincides with 
the absolute minimum of the energy, we have determined it by first looking for
the local minima ``visited'' by a sufficiently long trajectory at a fixed,
not-too-low, temperature value (tipically, $T=0.08$). More precisely, we
act as follows: we sample the trajectory every $\Delta t$ time units and
consider the configuration 
$(x_i(n \Delta t),y_i(n \Delta t),\dot x_i=0,\dot y_i = 0)$ as the
initial condition for the overdamped dynamics
\begin{equation}
{\dot x}_i = - \frac{1}{\gamma} \frac{\partial H}{\partial x_i}  \quad ,
\label{overdamp}
\end{equation}
where $\gamma$ is an irrelevant damping constant fixing the time scale
(an equivalent equation holds for $\dot y_i$). As a consequence, the 
phase point evolves towards the local minimum, the basin attraction of which 
contains the initial condition. Such a local minimum is the so-called 
``inherent state'' \cite{still2}. Upon repating this procedure over and over, 
we can construct a large ensemble of inherent minima: the minimal-energy state 
is eventually identified with the NC.

By comparing with \cite{irb_priv}, we have verified that, for the sequences
S1 and S3, this method allows for a correct identification not only 
of the NC but also of the 10 lowest energy configurations (with a few
differences due to the contribution of the harmonic potential $V_1$, replaced
by a rigid bond in the original model \cite{still}~). Considering that our 
method is rather straightforward in comparison to the quite elaborate 
Monte Carlo techniques implemented in \cite{irback2}, this is a first 
indication of the advantage of the dynamical approach. To be more precise, 
simulations for a time $t = 250,000$ (with sampling-time $\Delta t = 5$ 
and minimization-time $t=125$ -- $\gamma=2.5$) typically allow for a correct
identification of the NC and of the lowest energy minima with an accuracy of 
$10^{-5}-10^{-6}$ 
in energy and $10^{-2}$ for what concerns the angular distance $\delta$.
\begin{figure}[ht]
\cl{
\psfig{figure=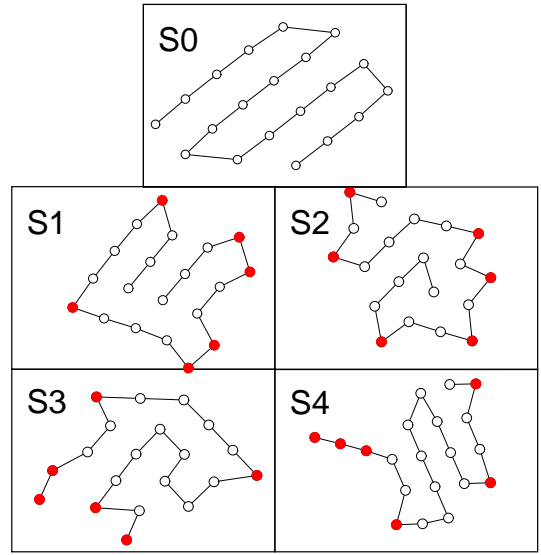,angle=-90,width=7truecm}
}
\caption{
The NCs corresponding to the five examined sequences. Full and  open
dots denote polar and hydrophobic residues, respectively.}
\label{nc}
\end{figure}

The distribution of polar residues is such that the formation of a stable 
hydrofobic core is possible in S1 and S2, while the concentration of the
residues at the ends in S3 and S4 induces the formation of fairly unstable
filaments.

In Table III we report the potential energy $V_{NC}$ of the native state
together with the energy gap $V_{gap}$ separating the NC from the second
lowest energy level. As expected the lowest energy corresponds to the
homopolymer sequence, while the other native-state energies are quite close
to each other. The largest gap is, instead, found for the sequence S1: it
is more than 3 times larger than the $V_{gap}$-value for S0 and S2 and
more than 30 times larger than in S3 and S4. According to the ``Shaknovich 
criterion" \cite{shak,shak1}, a protein with a large energy gap between
the NC and the first non-native (compact) configuration folds rapidly.
Therefore, one expects that S1 is a much faster folder, while S3 and 
S4 should really be slow folders. On the other hand, it has been recently 
shown that the folding dynamics depends on the whole energy landscape 
and not only on the energy gap \cite{orland}. In this sense, one should
consider that such criteria may provide useful guidelines for an approximate
identification of good folders.

\begin{table}[ht]
\caption[tabone]{
The minimal potential energy $V_{NC}$ (corresponding to the NC) is 
here reported together with the energy gap $V_{gap}$, the number of nearest
neighbours local minima of the NC, and the folding temperature $T_F$
for the 5 considered sequences.
}
\vskip 0.3 truecm
\begin{tabular}{rrrrrr}
\hline
\hfil  \hfil \hfil & \hfil S0 \hfil & \hfil S1 \hfil &
\hfil  S2  \hfil & \hfil S3  \hfil & \hfil S4 \hfil \hfil \\
\hline
$V_{NC}$ &  -7.0422  & -4.6666 & -5.1234 & -4.6283 & -4.6661 \\
$V_{gap}$ &  0.0255  & 0.0922 & 0.0244 & 0.0025 & 0.0017  \\
$n_{0}$ & 6  & 38 & 33  &  3 & 28  \\
$T_{F}$ & 0.036  & 0.065 & 0.048  &  0.036 & 0.046  \\
\hline
\end{tabular}
\end{table}

\subsection{Folding Temperature $T_F$}

A commonly used definition of the {\it folding temperature} $T_F$ (i.e.
of the temperature below which the polypeptide chain is predominantly in 
the native configuration) states that $T_F$ is the temperature for which 
the probability to visit the NC is $1/2$. At finite temperatures, in 
off-lattice simulations, the NC is never exactly met: this implies that
the implementation of the above definition requires defining a ``visit''
as a ``close-encounter'' with the ambiguity connected with the relative.
One could simply state that the NC 
has been ``visited'' whenever the phase point enters its basin of attraction.
In practice, we have verified that this is too narrow a criterion to be 
utilized for a meaningful definition of $T_F$. Accordingly, we have preferred 
to compute the probability $P(T)$ to visit the basin of attraction of either 
the NC or one of its neighbouring metastable states: in what follows we shall 
refer to this ensemble of attraction basins as ``native valley''. The 
definition of $T_F$ is, therefore, implicitely given by the constraint
$$
P(T_F) = 0.5  \quad .
$$
The metastable states have been identified by following the
sequence of minima visited during unfolding-dynamics simulations. In fact,
if the temperature $T$ is neither too small nor too high, a generic
trajectory explores the native valley jumping among all minima around 
the NC. As a result, we have observed that the number $n_0$ of minima
surrounding the NC is maximal for the sequence S1, while it reduces 
dramatically for S0 and S3 (see Table III for more details). This indicates
that many different pathways can lead to the folded structure in the case of 
S1, S2 and S4, while a few paths exist for S0 and S3.

\begin{figure}[ht]
\cl{
\epsfig{figure=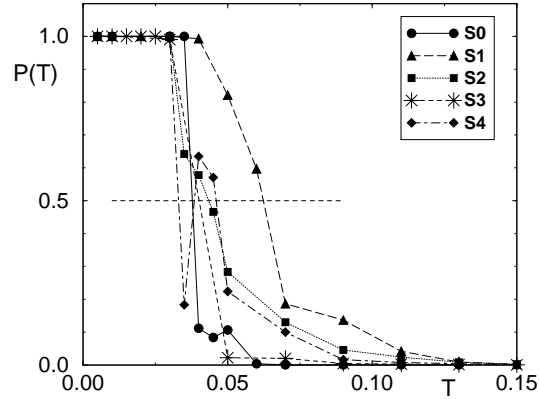,angle=-90,width=7truecm}
}
\vskip 0.5 cm
\caption{
The probability $P(T)$ versus the temperature $T$ in lin-log scales. The 
measurements have been performed during US's of duration $t=250,000$, where 
an overdamped relaxation scheme has been applied 
every $\Delta t=5$ to find the underlying local minimum.
}
\label{f2}
\end{figure}

The probability $P(T)$ to be in either the NC or one of its $n_0$ 
neighbours has been measured at different temperatures for all the sequences.
The data reported in Fig.~\ref{f2} reveal quite an abrupt decrease of $P(T)$
for both S0 and S3, while a smoother behaviour has been found for the three
other sequences. 
The corresponding folding temperatures are reported in Table III. The highest 
value is found for S1 ($T_F=0.065$),  while the lowest for S0 ($T_F=0.036$). 
In the next section we will show that $T_F$ coincides with the minimal 
temperature for which the NC is dynamically accessible within a ``realistic'' 
lapse of time.

\subsection{Energy Funnel}

\begin{figure}[ht]
\cl{
\epsfig{figure=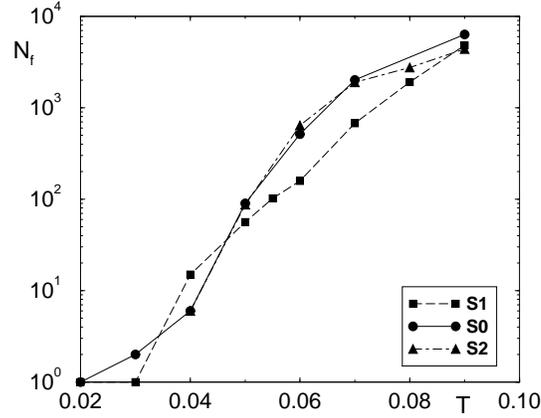,angle=-90,width=7truecm}
}
\vskip 0.5 cm
\caption{Number of distinct minima obtained during a US of duration $t=12,500$, with
an overdamped relaxation scheme applied every $\Delta t=0.05$ to locate the 
nearest inherent minima.
}
\label{f1}
\end{figure}

The energy landscape associated to three of the five mentioned sequences has 
been investigated by examining the distribution of local minima obtained by 
performing US's at various temperatures. For each value of $T$, we have made
a long simulation of duration $t_f=12500$, applying the above described
overdamped integration scheme every $\Delta t = 0.05$ time units, to
identify the inherent minima. As a result, we have been able to determine
the number $N_d(t,T)$ of distinct minima visited up to time $t$ at temperature
$T$. For each sequence, we have found that $N_f = N_d(t_f,T)$ decreases 
noticeably with $T$ 
(see Fig.~\ref{f1}, where we have reported the results for the sequences S0, 
S1 and S2). This observation, analogous to what recently found for a 
$\beta$-heptapeptide in methanol solution \cite{daura}, is quite obvious, 
since at low temperatures the trajectory remains trapped into local minima. 
In particular, we see that for the sequence S1, 6324 different minima have 
been identified at $T=0.09$ while only 6 distinct minima have been visited 
at $T=0.04$.

\noindent
A detailed investigation would require considering the hidden dependence
of $N_d$ on the integration time $t_f$. The practical computer-time
limitations obliged us to study one case only. In Fig.~\ref{f1} we observe 
that the sequence S1 is characterized by an almost exponential increase of 
$N_f$ with temperature. This confirms that the native valley is well 
connected to many other valleys of the energy landscape for the S1 sequence, 
while much fewer connections are found for S2 and S0.

\begin{figure}[ht]
\cl{
\epsfig{figure=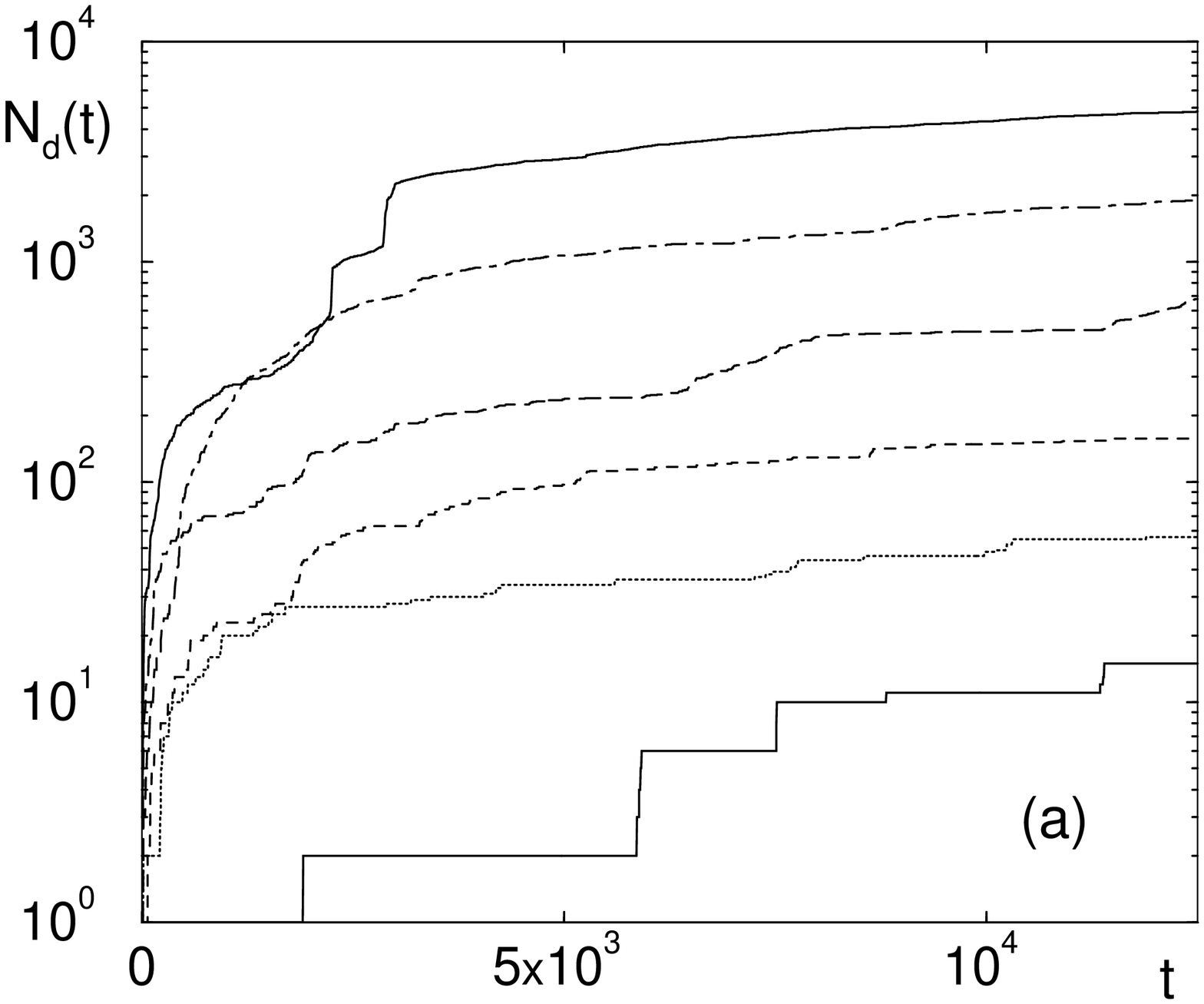,angle=-90,width=5.5truecm}
\epsfig{figure=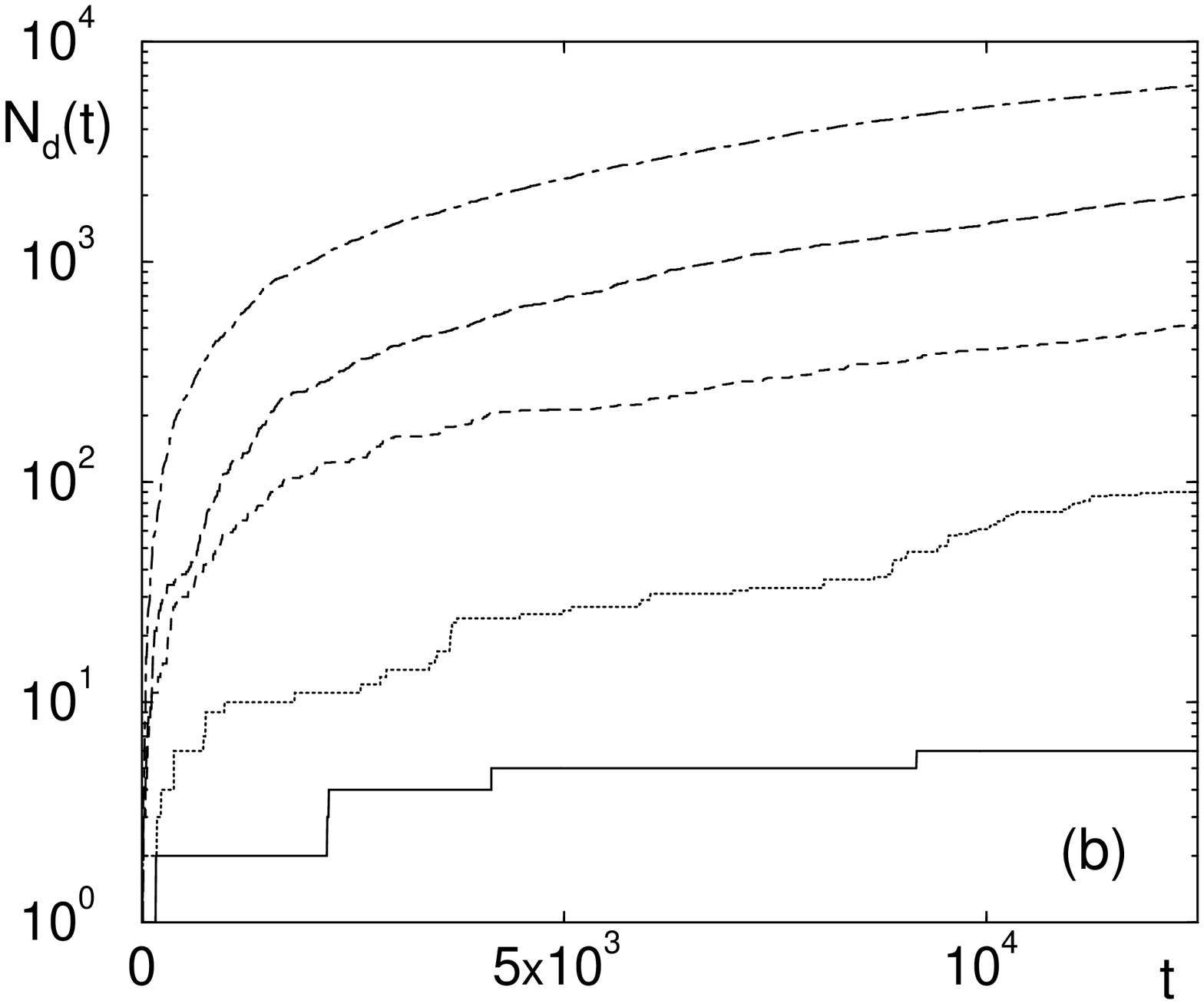,angle=-90,width=5.5truecm}
}
\vskip 0.5 cm
\caption{
Number of distinct inherent minima $N_d(t,T)$ visited during a US of
duration $t=12500$. Panel (a) contains the results for the S1 sequence:
from bottom to top the temperatures are $T = 0.04$, 0.05, 0.06, 0.07, 0.08, 
and 0.09; panel (b) refers to the sequence S0: from bottom to top the 
temperatures are $T = 0.04$, 0.05, 0.06, 0.07, and 0.09.}
\label{ndt}
\end{figure}

A further interesting quantity to study is the temporal growth of the number 
of distinct inherent minima visited. In Fig.~\ref{ndt}, we have reported 
$N_d(t,T)$ at various temperatures for the sequences S1 and S0. It is evident 
that the heteropolymer, at low temperatures, is trapped in the native valley 
for the simulation duration, while for $T >T_F$, it starts visiting 
other valleys. 

The nonuniform growth exhibited by $N_d$ (see, for instance, the highest
temperature simulations for S1) are suggestive of the existence of
different valleys: as soon as some specific ``passes'' are overcome, a
new landscape appears making new valleys easily accessible.

\begin{figure}[h]
\cl{
\epsfig{figure=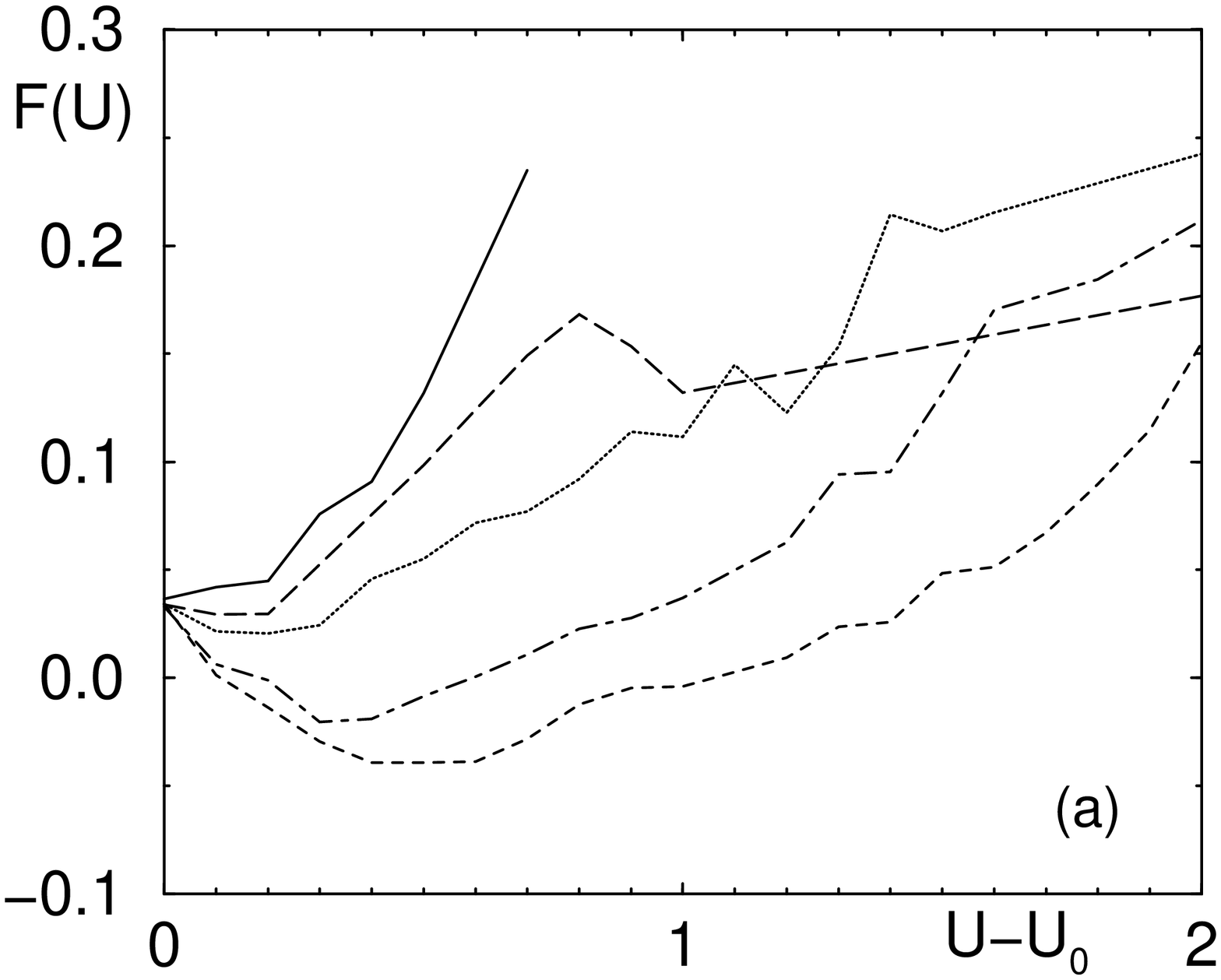,angle=-90,width=5.5truecm}
\epsfig{figure=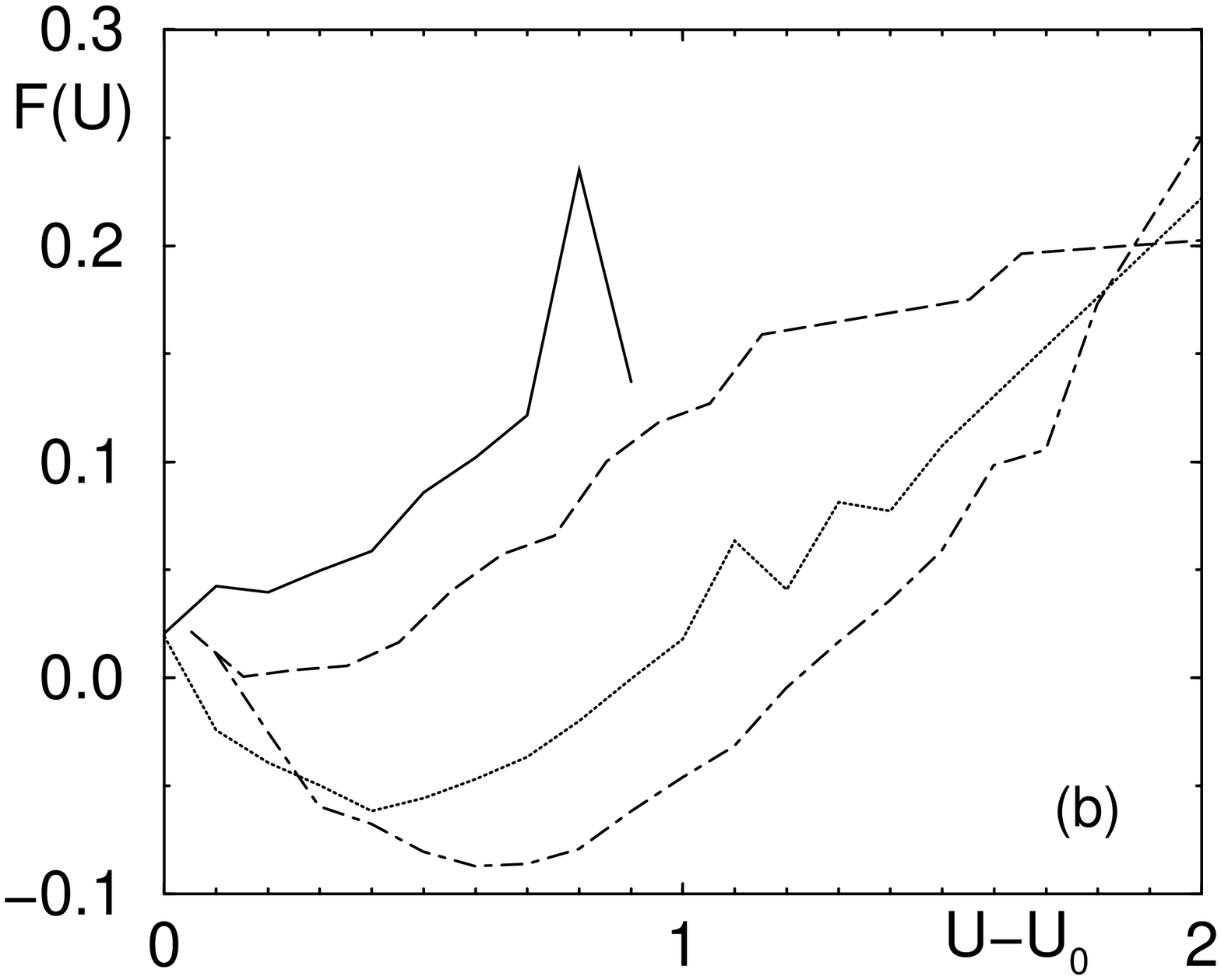,angle=-90,width=5.5truecm}
}
\vskip 0.5 cm
\caption{
Free energy $F(U)$ as a function of $U-U_0$ (where $U_0$ is the energy of the 
NC) for various temperatures. Panel (a) refers to the S1 sequence with 
temperatures $T = 0.05$, 0.06, 0.07, 0.08, and 0.09 from top to bottom;
panel (b) refers to the S0 sequence with temperatures $T = 0.05$, 0.06, 0.07, 
and 0.09, from top to bottom. The origin of the free energy is fixed 
arbitrarily.}
\label{free}
\end{figure}

In order to give a pictorial description of the energy landscape, we have
decided to define the ``free energy'' 
\begin{equation}
F(U) = -T \ln[Q(U,T)]
\label{freeq}
\end{equation}
where $Q(U,T)$ is the probability that, at temperature $T$, the heteropolymer 
is in one of the inherent minima whose energy lies within the interval 
$[U,U+\delta U]$.
The results for $S1$ and $S0$ are reported in Fig.~\ref{free}, where
$Q(U,T)$ has been determined by fixing $\delta U = 0.1$~. It is evident
that at sufficiently low temperatures, the minimum of $F(U)$ is located at
the energy $U=U_0$ of the NC, while at higher temperatures, the minimum
shifts to larger values, indicating that the NC is no longer the favoured 
thermodynamical state.  For the S1 sequence, this occurs for $T > 0.07$, while
for S0, the NC is no longer favoured already for $T = 0.05$. These numbers 
are compatibile with the previously estimated values of the folding 
temperature and suggest an alternative definition of $T_F$ pointing more 
directly to the folding process as to a phase-transition (with all limitations 
due to the fact we are referring to finite systems).

\section{Dynamics}

In the previous sections we have adopted the widespread attitude of
describing the folding process with concepts and tools borrowed from
the language of equilibrium thermodynamics. This approach proves partially
effective in singling out differences between good and bad folders, although 
the heuristic criteria proposed so far reveal some degree of ambiguity. 
However, the folding process can be viewed as a transient evolution 
towards a uniquely selected native state. In this perspective, a dynamical 
description of the folding process seems more appropriate than a purely 
thermodynamic one. First, by looking at the evolution, one can identify the 
accessible pathways towards the native configuration and thereby estimate the 
folding time. Furthermore, the relative length of the proteins makes the use
of thermodynamical concepts rather questionable.

\begin{figure}[h]
\cl{
\psfig{figure=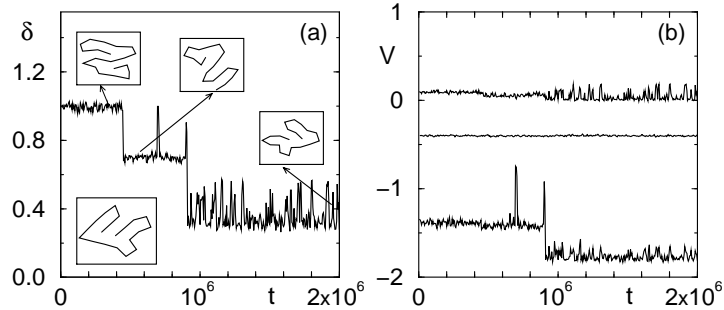,angle=-90,width=10.truecm}
}
\vskip 0.5 cm
\caption{A typical FS for the sequence S1 at $T=0.06$. (a) The distance
$\delta$ with respect to the NC is plotted versus time. The three upper insets
contain snapshots of the configuration in each of the three plateaus (the
NC is reported in the lower-left inset for comparison). The three component
of the potential energy $V_2$, $V_1$, and $V_3$ are reported in panel (b)
(from top to bottom) for the same FS. The potential energies are arbitrarily 
shifted along the vertical axis.}
\label{d81}
\end{figure}

A proper observable to look at is the angular distance $\delta (t)$, defined 
in Sec.~3. In Fig.~\ref{d81}(a), one can notice how, for $T_G < T < T_F$, the 
approach to the native valley ($\delta \approx 0$) of S1 is characterized by 
a series of jumps that correspond to successive rearrangements of the chain 
configuration. This process can be better appreciated by looking at the 
snapshots taken in the various plateaus (the NC is reported for the sake of 
comparison). Notice that although the ``asymptotic'' average value $\delta_a$ 
of the distance is not numerically too small ($\delta_a \simeq 0.3$), the
dynamical configuration looks very similar to the NC (within a left-right
symmetry transformation that in 2D has to be always considered). In other 
words $\delta_a$ is quite sensitive an indicator.

In Fig. \ref{d81}(b) one can look at the evolution of the three components 
of the potential energy (see Eqs.~(\ref{v1})) for the same trajectory 
as in Fig. \ref{d81}(a). The harmonic component $V_1$ is, as expected, 
completely 
insensitive to the various structure changes, while the angular energy $V_2$ 
limits itself to displaying larger fluctuations in the asymptotic regime. 
The only interesting behaviour is exhibited by the long-range potential 
energy $V_3$, which closely reproduces the jumps of $\delta(t)$ with the
only exception of the first one. 

The time needed for S1 to enter the native valley is tipically on the order
of ${\cal O}(10^6-10^7)$ adimensional units. In physical units, this
corresponds to ${\cal O}(10^{-5})$ seconds, a much shorter time than that 
typically observed in real proteins\footnote{A rough estimate of the ``real'' 
time scale involved in the folding process described by our model can be 
derived from the period of small oscillations 
$\tau_{LJ} \sim \sqrt{m \sigma^2/\varepsilon}$ in the Lennard-Jones potential, 
where $m$ is the typical mass of an aminoacid, $\varepsilon$ is the depth of 
the potential well and $\sigma$ is the equilibrium distance. In our model, 
$m=1$~, $\sigma \sim 1 $ and $\varepsilon=1$,
while for an aminoacid $m \sim 1-3 \cdot 10^{-22}$ g., the equilibrium
distance is $7-9 \cdot 10^{-8}$ cm. and the enregy of a hydrophobic 
interaction is $\sim 1-2 \cdot 10^{-13}$ erg. Therefore, 
$\tau_{LJ} \sim 4-6 \cdot 10^{-12}$ sec., so that the folding time-scale for 
the sequence S1 is ${\cal O} (10^{-5})$ secs.}. 
It seems reasonable to conjecture that the two-dimensional character of the
space is the main responsible for the shortness of the time scale. Indeed, 
the more stringent geometrical constraints induce a faster folding in 2D than 
in 3D. Moreover, the relatively small chain length ($L=20$) contributes to 
fastening the folding process, as well.

Nonetheless, it is instructive to point out that the average time scale of 
the folding process is already six orders of magnitude larger than the typical
scales of equilibrium vibrations: this testifies to the meaninfulness of
the model itself that is likely to be strenghtened by extending it to 3D.
 
The behaviour of the other sequences exhibit both differences and analogies. 
In order to observe a convincing convergence to the NC, one has to consider 
smaller temperatures. From a biological point of view, the energy scale is
very important, as a meaningful protein has to fold in a specific temperature
range. On the other hand, as the energy  scale is somehow arbitrary in our
model, one would like to understand whether the difference between the
various sequences reduces to fixing the temperatures at which specific 
phenomena are observed. 

\begin{figure}[h]
\cl{
\epsfig{figure=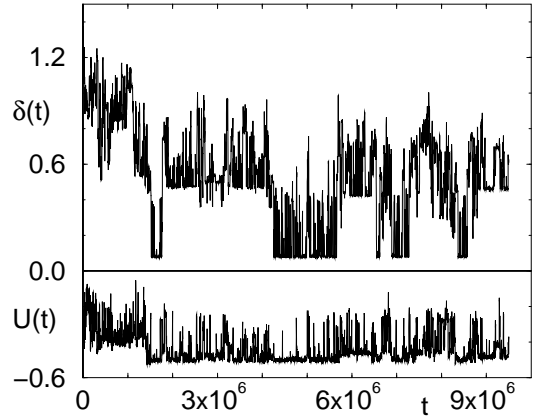,angle=-90,width=7truecm}
}
\vskip 0.5 cm
\caption{A typical FS for the sequence S0 at $T=0.04$.
The distance $\delta$ with respect to the NC and the potential energy $U$ are 
plotted versus time. The potential energy $U$ is arbitrarly shifted along 
the vertical axis.}
\label{d00}
\end{figure}

\begin{figure}[h]
\cl{
\epsfig{figure=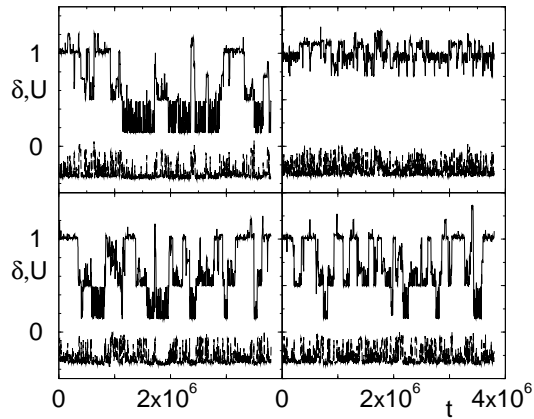,angle=-90,width=7truecm}
}
\vskip 0.5 cm
\caption{Four samples of FSs for the sequence S3 at $T=0.04$.
The distance $\delta$ with respect to the NC (upper curves) and the 
overall potential energy $U$ (lower curves) are plotted versus time. 
The potential energy $U$ is arbitrarly shifted along the vertical axis.}
\label{d50}
\end{figure}

The data reported in Figs.~\ref{d00} and \ref{d50} refer to S0 and S3, 
respectively. The evolution of $\delta$ and $U$ for $T=0.04$ is qualitatively
similar to that exhibited by S1 at temperature $T=0.06$. Nevertheless,
there is an important difference that may have relevant biological
implications: the fluctuations of $\delta$ (and, similarly, of the total
energy $U$) within what appears to be the native valley are larger for
S0 and S3 than for S1 (the standard deviation is approximately equal to 0.14,
0.12, and 0.08, in the three cases), even though the temperature is comparably 
smaller than in the latter case. This points to a more clear configurational
stability of the folded state of S1.

\section{Concluding remarks}

Five different sequences of a 2d toy-model of aminoacid chains 
have been studied in detail. Time averages of thermodynamical indicators 
have been analyzed in order to check their consistency in 
discriminating between one specimen of ``good folder'' (S1)
and a set of four very different sequences. These equilibrium simulations
yield controversial conclusions: the various indicators often attribute 
different rankings to the various sequences (in one case, S1 is even located 
among the worst candidates for a ``good folder''). 

A clear identification of S1 as a ``good folder'' is provided by the 
foldability factor $Q$ and by the Camacho, Klimov and Thirumalai criterion 
applied to $\sigma^*$. The Shaknovich criterion too points in this 
direction, as S1 exhibits the largest energy gap $V_{gap}$.

In summary, applying a ``majority rule'' we are led to conclude
that S1 is actually a ``good folder''. Nonetheless, it would be
definitely better to rely upon less ambiguous criteria for achieving such 
a conclusion. The direct inspection of the dynamics of the sequences
provides a more clear scenario. Simulations performed at temperatures close
to $T_F$, reveal that an initial ``coil'' state evolves towards its native 
configuration entering the energy funnel by a sequence of configurational 
jumps. In order to obtain a similar scenario for the other sequences, 
the temperature has to be significantly lowered and this is not sufficient,
since the relative fluctuations within the native valley are comparably
larger.

The analysis of the structure of the energy landscape provides complementary 
indications that are consistent with the dynamical description. In particular,
the ``free-energy'' $F(U)$ defined in Eq.~(\ref{freeq}) shows that the NC
is a minimum of $F(U)$ only below a temperature close to $T_F$ (defined
either by looking at the behaviour of $\delta$ or $\chi$). Above $T_F$, the 
minimum shifts away, suggesting that the stable thermodynamical state differs 
from NC. Moreover we have discovered that in $S1$ the ten ineherent minima 
of lowest energy are dynamically connected to the NC (i.e. all of them 
belong to the native valley), while in S0 and S3 only a few are directly 
connected to the corresponding NC's. This indicates that the gap height,
although certainly important, does not provide a full account of the
relevant folding properties. Preliminary investigations \cite{tiberio} suggest 
that such an information must be complemented with the ``connectivity'' 
between the NC and the other low-energy minima and with the height and 
``shape'' of the barriers separating the NC from the inherent minima. 

\acknowledgements
We warmly acknowledge the active contribution of Annalisa Tiberio to the 
present work and the useful interaction with Anders Irb{\"a}ck.
Part of this work was performed at the Institute of Scientific Interchange
in Torino, during the workshops on `` Complexity and Chaos '', 
June 1998 and October 1999.  We acknowledge CINECA in Bologna and INFM 
for providing us access to the parallel computer CRAY T3E under the  
grant ``Iniziativa Calcolo Parallelo''.


\end{article}
\end{document}